# Forecasting influenza hospital admissions within English sub-regions using hierarchical generalised additive models

Jonathon Mellor[1*], Rachel Christie[1], Christopher E Overton[1,2], Robert S Paton[1], Rhianna Leslie[1], Maria Tang[1], Sarah Deeny[1], Thomas Ward[1]

1. UK Health Security Agency, Data Analytics and Science, Noble House, London, United Kingdom
2. University of Liverpool, Department of Mathematical Sciences, Liverpool, United Kingdom

*Corresponding Author: Jonathon.Mellor@UKHSA.gov.uk

## Abstract

**Background**
Seasonal influenza causes a substantial burden on healthcare services over the winter period when these systems are already under pressure. Policies during the COVID-19 pandemic supressed the transmission of season influenza, making the timing and magnitude of a potential resurgence difficult to predict.

**Methods**
We developed a hierarchical generalised additive model (GAM) for the short-term forecasting of hospital admissions with a positive test for the influenza virus sub-regionally across England. The model incorporates a multi-level structure of spatio-temporal splines, weekly seasonality, and spatial correlation. Using multiple performance metrics including interval score, coverage, bias, and median absolute error, the predictive performance is evaluated for the 2022/23 seasonal wave. Performance is measured against an autoregressive integrated moving average (ARIMA) time series model.

**Results**

The GAM method outperformed the ARIMA model across scoring rules at both high and low-level geographies, and across the different phases of the epidemic wave including the turning point. The performance of the GAM with a 14-day forecast horizon was comparable in error to the ARIMA at 7 days. The performance of the GAM is found to be most sensitive to the flexibility of the smoothing function that measures the national epidemic trend.

**Interpretation**

This study introduces a novel approach to short-term forecasting of hospital admissions with influenza using hierarchical, spatial, and temporal components. The model is data-driven and practical to deploy using information realistically available at time of prediction, addressing key limitations of epidemic forecasting approaches. This model was used across the winter for healthcare operational planning by the UK Health Security Agency and the National Health Service in England.

## 1. Introduction

Seasonal influenza is a substantial winter burden on the primary and secondary healthcare system in England, causing mortality and morbidity for patients [1, 2, 3, 4, 5, 6]. The most recent winters of 2020/2021 and 2021/2022 being notable exceptions, with little influenza in circulation, likely due to the impact of COVID-19 targeted non-pharmaceutical interventions. Influenza admissions [7] reduce the capacity of the healthcare service to provide emergency care and elective treatment. It is therefore essential for healthcare operational planning to be able to understand the expected demand on the system.

In the United Kingdom, influenza leads to the hospitalisation of around 28,000 people annually, with significant variation between seasons [8]. The disease disproportionately hospitalises young children, the elderly, and adults with high-risk comorbidities [9]. Between the 1996 – 2009 seasons, weekly population hospitalisation rates in the UK were highest in adults over 75 (252 per 100,000), adults aged 65-74 (101 per 100,000) and children under 5 (93 per 100,000) [8]. Being able to forecast hospital admissions with influenza accurately allows administrators to deploy mitigations faster, reducing pressures. In winter 2022/23, the uncertainty around the likely trajectory of the influenza season, and the impact of COVID-19 waves on system resources, increased the need for fast, flexible models that could be used when population level epidemiological parameters (such as immunity and social mixing behaviour) are unknown. Short term forecasts of influenza hospital cases would allow public health officials and health system leaders to make informed decisions around policy and response during the season, such as the deployment of staff, mutual aid between regions or hospitals, changes to triage or discharge practice, or other methods to increase bed capacity. Influenza forecasting is a well-studied field, with a wide range of methodologies applied [10, 11]. Previous work includes a mix of statistical and mechanistic approaches, with the aim of predicting the timing, location, and magnitude of epidemic waves. These include when peaks in infection and hospitalisations occur [12, 13], how large the peaks will be [12], and incorporating leading indicators [14]. The start of an outbreak is also of interest and can be forecast at fine spatial scales [15]. In the US, many of these policy-relevant metrics are predicted, for example in the Centre for Disease Control (CDC) FluSight forecasting challenge where a range of models are ensembled [16]. In winter 2022/2023, several factors (uncertainty around population dynamics, the early season, and lack of comparable seasons to train models) meant that a less mechanistic model was of increased operational utility.

In this paper we present a highly generalisable method for short-term admissions forecasts, which could be applicable to a range of infectious diseases and in a range of settings. It was deployed by UK Health Security Agency in winter 2022/23. The method could be used nationally, and within local healthcare regions using data within the organisations and publicly available reference information, without the need for high performance computational hardware.

The model is a hierarchical generalised additive structure with a spatial component to extrapolate forward two weeks the national, regional, and sub-regional trends within England. We score this model's performance by a range of metrics and contrast it with a commonly used time series forecasting approach, highlighting its utility in public health response operations.

## 2. Methodology

### 2.1 Data

An NHS Trust is an organisational unit within the National Health Service (NHS) serving a geographical area or a specialised function; there are 137 acute (containing emergency departments) Trusts in England. Each Trust is within a Sustainability and Transformation Partnership (STP) grouping, formed for NHS Trusts to develop plans to transform healthcare delivery in their shared area [17], of which there are 44 in England. Each STP has an associated boundary, each within one of the seven NHS England commissioning regions.

#### 2.1.1 NHSE SitRep

NHS England (NHSE) data is provided by individual NHS acute trusts in England, who deliver a daily situation report (SitRep) covering the previous 24 hours on urgent and emergency care activity (UEC) by 11am each day [18]. The reporting process for many trusts is automated and completed via web form [19]. The data records both hospital bed occupancy and new patients in the past 24 hours with a laboratory-confirmed positive influenza test [20]. We assume positive test patients in the last 24 hours are analogous to admissions throughout this paper, though they likely also include hospital acquired infections. Linking information in the data includes the Trust, STP [17], and NHS region. The data available for this study covers 01 Dec 2021 to the week ending 29 Jan 2023.

The SitRep is used to forecast influenza admissions as it is timely and near complete. The SitRep is provided weekly allowing for analysis in real time as the epidemic progresses and is used by the NHS for operational planning. As of 05 Jan 2023, 137 NHS Trusts were reporting data to the SitRep, of which 123 reported data on more than 99% of days. This provides high coverage of the English population and is representative in all regions. Equally, this allows modelling at NHS Trust or STP level, increasing the operational utility of the approach. Furthermore, patients must have laboratory-confirmed positive influenza tests to be included in the SitRep, which increases the reliability of reporting across NHS Trusts, and avoids convoluting counts with other influenza-like-illnesses. Though a high coverage and well specified data source, there are several Trusts which misreported admissions counts due to the rapid turnaround of the reporting and the fact this is the first reemergent influenza season since the collection started. This can be handled by data cleaning during the season, though it can be challenging to discern between true and false reporting, and partially accounted for in modelling. Trusts that started reporting part way through the season or had clear misreporting were excluded from the analysis.

#### 2.1.2 STP boundaries

When modelling trends in spatial data we expect nearby locations to be closely related. In this case, we expect neighbouring STPs to be correlated. We create a network based on STP boundary adjacency, such that touching neighbours are connected and weighted by the distance from centroids. This was done such that a Markov Random Field [21] spatial smoothing approach could be included within the modelling.

STP boundary files were sourced from the ONS Open Geography Portal [22]. These contain digital vector boundaries, or polygons, for STPs in England as of April 2021. The STP structure of England, NHS commissioning regions, individual Trust locations and catchment population sizes are shown in Supplementary Figure A.

### *2.1.3 Respiratory Local Authority to Trust Mapping*

Larger STPs will service correspondingly large population catchments, and therefore have higher daily counts of influenza patients. We normalise admissions by population catchment for an STP, giving a per capita rate. As the hospital a patient visits depends on location, choice [23] and age group [24] we create a probabilistic mapping between lower tier local authority (LTLA) administrative regions and NHS Trusts, linking populations and hospitals. This mapping uses counts of patient records from Secondary Use Services – All Patient Care [25], linking a discharge location (LTLA) to the service provider (NHS Trust), in a similar manner to the *covid19.nhs.data* R package [26]. The counts include all patients admitted from 01 Jan 2021 to 11 Dec 2022 with a respiratory-related primary or secondary diagnosis code, excluding COVID-19 patients – using COVID-19 test linkage with the Second Generation Surveillance Service (SGSS) [27]. ONS 2019 local authority population estimates are used to produce a weighted sum of populations across trusts of their feeder LTLAs, giving an effective population catchment size. Since the data were modelled at STP level for these forecasts, admissions and populations were aggregated from reporting Trusts within an STP to calculate the rate.

### *2.2 Model*

### *2.2.1 Generalised additive model structure*

To forecast hospital admissions with influenza, we first assume a semi-mechanistic model, where the growth rate of the epidemic continues forward in time from the most recently available data. We assume in an exponentially growing epidemic we have the per-capita rate, with population $p$, of hospitalisations at time $t$, $H(t)/p$,

$$\frac{H(t)}{p} = \frac{H(0)}{p} e^{rt}.$$

Instead of assuming a constant growth rate $r$, we can consider the growth as a smooth function of time $s(t)$, i.e.

$$\frac{H(t)}{p} = \frac{H(0)}{p} e^{s(t)}.$$

Such a model can be used to generate short-term forecasts by assuming that the growth rate remains constant outside of the data window, i.e. $s(t) = r * t$ when $t \geq t_{\max}$, where $t_{\max}$ is the final date for which we have data. To fit such a model, we need to estimate the smooth function $s(t)$ for $t \in [0, t_{\max}]$. Taking the logarithm of both sides, this becomes

$$\frac{H(t)}{p} = \frac{H(0)}{p} e^{s(t)} \Rightarrow \log\left(\frac{H(t)}{p}\right) = \log\left(\frac{H(0)}{p}\right) + s(t).$$

$$\Rightarrow \log(H(t)) = \log\left(\frac{H(0)}{p}\right) + s(t) + \log(p).$$

The number of hospital admissions on each day, $H(t)$, is an overdispersed integer valued count, which can be considered as a random sample from a negative binomial distribution. Therefore, the smoother $s(t)$ may be estimated as $y(t)$ by using a GAM with negative binomial error structure, with an intercept term $\beta_0$, smoothing function of time $f(t)$, and offset for the population size, i.e.

$$\log(y(t)) = \beta_0 + f(t) + \log(p).$$

### *2.2.2 Incorporating STP level structures and population size*

We aggregate the data to STP level, so all sub-regions have at least one reporting hospital, which allows us to create a connected network of adjacent sub-regions. We assume a seasonal influenza outbreak will have a national epidemic curve component, as well as trends sub-nationally (STP level), but these sub-national trends are not fully independent of one another in shape. A hierarchical GAM is used to account for the group-level structure. Following the "GS" form from Pedersen [28], we add a global smoother (national level) and group-level smoothers (STP level) that have the same wiggliness – a "**G**lobal smoother with individual effects that have a **S**hared penalty" [28]. To further account for between-group correlation of adjacent STPs, we estimate an intercept using spatial smoothing, correlating adjacent STPs using a Markov Random Field (MRF) approach and a network based on adjacency in Section 2.1.3.

Each STP has different population catchments, so we include an offset, $p_i$, to calculate an admissions rate per capita for a given STP $i$; for reporting we transform the forecasts back to counts for ease of comparison. We include a regional random effect $\delta_{region_i}$. As hospital admissions and test reporting are impacted by the day of week (DOW), we include a random effect $\delta_{d_t, stp_i, region_i}$ of DOW by STP nested within region. As reporting quality varies between STP and region the DOW random effect is fitted for each STP, and for those STPs with limited information on effect size, this term converges to the regional effect. The STP intercept is approximated by the Markov Random Field $f_{mrf}(i)$, the STP temporal trends by the individual effect shared penalty $f_{stp_i}(t)$ and a national trend by $f_{nat}(t)$. This gives the model formulation as

$$log(y_i(t)) = \beta_0 + f_{nat}(t) + f_{stp_i}(t) + f_{mrf}(i) + \delta_{d_t, stp_i, region_i} + \delta_{region_i} + \log(p_i).$$

Forecasts at multiple geographies are policy relevant: individual NHS STPs, regions, and nationally. To obtain these, we aggregate the individual time series forecasts using a "bottom up" approach, where disaggregate predictions are aggregated to higher levels, such as STP to region and to nation. This is often more accurate when disaggregate data is available than "top down" methods, where data is first aggregated to a high level, forecasts generated, then disaggregated proportionally to the lower levels [29] [30]. Where STP level count data is weak, the model is still accounting for national trends, which partially mitigates low data quality in individual STPs.

### 2.2.3 Implementing the model

We implement the GAM using the R package *mgcv* [31]. In *mgcv*, the negative binomial distribution is parameterised in the terms of the mean, $\mu$, and variance, $\sigma^2$, such that $\mu = E[y]$ and $\sigma^2 = E[y] + \frac{E[y]^2}{\theta}$. In this formulation, the $\theta$ parameter is fitted by the package rather than specified. Thin plate splines are selected as the basis functions, with the optimum number selected by scoring different combinations, shown in Figure 5. Sensitivity analysis was conducted to understand the optimal model structure of the different additive components and hierarchical structures during model development. In *mgcv*, the parameter $k$ sets the dimensionality of the spline and so the number of basis functions is given by $k - 1$. We define the number of days per basis function as $t_d$ and the number of days the model is fit with as $t_{length}$. We round down the ratio of $t_{length}$ to $t_d$ to the closest integer to calculate $k - 1, i.e.$

$$k - 1 = \left\lfloor \frac{t_{length}}{t_d} \right\rfloor.$$

In this analysis $t_{length}$ is chosen as nine weeks, 63 days and $t_d \in [1,14]$ for both $f_{nat}(t)$ and $f_{stp_i}(t)$. The optimal model is found to be $t_d = 5$, so all results are shown for this.

### 2.2.4 Baseline ARIMA model

To contextualise the performance of our novel hierarchical GAM model we compare it with a commonly used approach. As a baseline, we chose an autoregressive integrated moving average (ARIMA) model [32], a widely employed technique in time series forecasting in epidemiology [33] [10] [34]. The structure of an ARIMA is explained in further detail by Nau [35], though we provide a summary of relevant parameters. A seasonal component is also incorporated, SARMIA, to capture periodic trends. $(S)ARIMA(p,d,q)[P,D,Q]_s$ describes the model, where $p$ is the number of autoregressive terms, past values of the time series; $d$ is the number of differences of the time series to make the series stationary; and $q$ is the number of lagged forecast errors included. $P, D, Q$ and $s$ stand respectively for seasonal autoregression, seasonal integration, seasonal moving average, and season period length. We transformed the influenza admission counts using $\log(x + 1)$ to account for the non-negative, exponential trends of the epidemic. We apply the ARIMA models implementing the Hyndman–Khandakar algorithm within the R package *forecasts* [36] via *fable* [37] which determine these parameters by minimising the AIC after differencing the data. As with the GAM, we use nine weeks of historic data to predict a two-week trend. Regional and national level trends are fit separately rather than aggregating the STP forecasts in a bottom-up approach. This differs from the GAM structure as we found that the non-hierarchical structure performed better for the ARIMA model; the scores for the hierarchical and non-hierarchical model can be found in the Supplementary Table A.

### 2.2.5 Model evaluation

The quality of each model is assessed by quantifying the calibration, sharpness, over and underprediction, bias, and median absolute error (MAE) of the predictions. Calibration is defined as the statistical consistency of the forecasts and true observations, for which we chose 50% and 90% interval coverage. Bias, as well as overprediction and underprediction, refers to if the predictions are higher or lower than the true values. The (Weighted) interval score was chosen as it combines over and underprediction, along with sharpness, a measure of the model's concentration of the predictive distributions [38] [39]. The MAE provides a more intuitive number to explain to end users of the forecast as it is in the natural units of the data shown. Scoring metrics were implemented with the *scoringutils* R package [39]. We forecast 14 days into the future and evaluate weekly predictions to replicate a real-world application. The performance is examined from the start to end of the winter 2022/23 seasonal epidemic wave. The predictive performance at STP, regional and national levels are examined for the $t \in [t_{max} + 1, t_{max} + 14]$ forecast horizon, as policy decisions are made at these higher geographies. Where appropriate, we score the predictions at $t_{max} + 7$ and $t_{max} + 14$ to highlight differences in the length of the forecast horizon. The MAE and bias are reported over time for each prediction to examine the trend across the influenza season.

## 3. Results

### *3.1 Influenza Season 2022/23*

The 2022/23 influenza season saw hospital admissions in England start to increase rapidly around the start of November and reach a peak by the end of December, after which the number of cases admitted to hospital decreased rapidly, shown in Figure 1. The data from the NHSE UEC SitRep suggest that the peak would have occurred between the 21 Dec 2022 and 29 Dec 2022 at approximately 1,000 new patients per day, a period of atypical social mixing patterns. However, there were likely data reporting issues and changes to triage behaviour caused by bank holidays, which could have contributed to the decrease in cases reported within that period. From an operational perspective the timing of this peak, just before most surveillance systems and teams were at reduced capacity makes this wave a particular challenge to forecast, as data is less timely and fewer complimentary data sources are available to validate trends against. Additionally, the day of week effect observed in the data was compounded by the timing of bank holidays, for example, Christmas day fell on a Sunday, the day of the week with lowest expected admissions – muddying the trends. The individual STP trends are shown in Supplementary Figures D – K, highlighting the variation in size of wave and consistency of reporting.

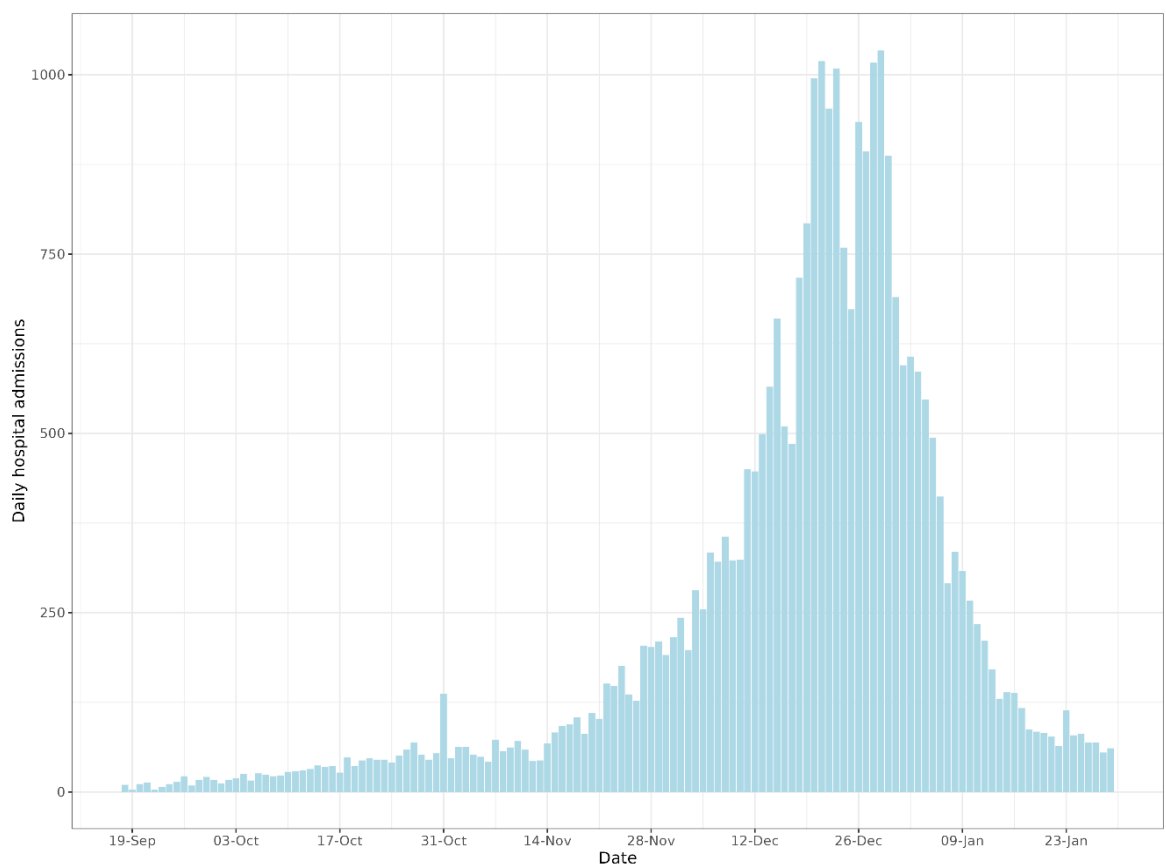

*Figure 1. National influenza 2022/23 season for England between 18 Sept 2022 and 29 Jan 2022, using the NHSE UEC SitRep.*

*3.2 Weekly Forecasts*

The national forecasts of the GAM and ARIMA models are shown in Figure 2, for the 2022/23 season in weekly intervals, to illustrate performance at different phases of the epidemic wave. The national forecast shows that the GAM is quicker to adapt to the change in direction of the epidemic wave, as shown by the forecasts starting 19 Dec 2022 to 02 Jan 2023. The ARIMA model continues the previously observed trend, as expected from an autoregressive model, causing it to overshoot at the peak, and overpredict on the subsequence downturn weeks, though the ARIMA performance appears strong in the initial growth phase. The prediction intervals for the GAM are generally wider than the ARIMA, capturing more overdispersion and variation in weekly reporting patterns. The central estimate near peak levels is closer to the data in the GAM model, with smaller underpredictions in the growth phase, and smaller overprediction in the decline, despite this also being an autoregressive model. Individual STP level forecasts are given in Supplementary Figures L – R.

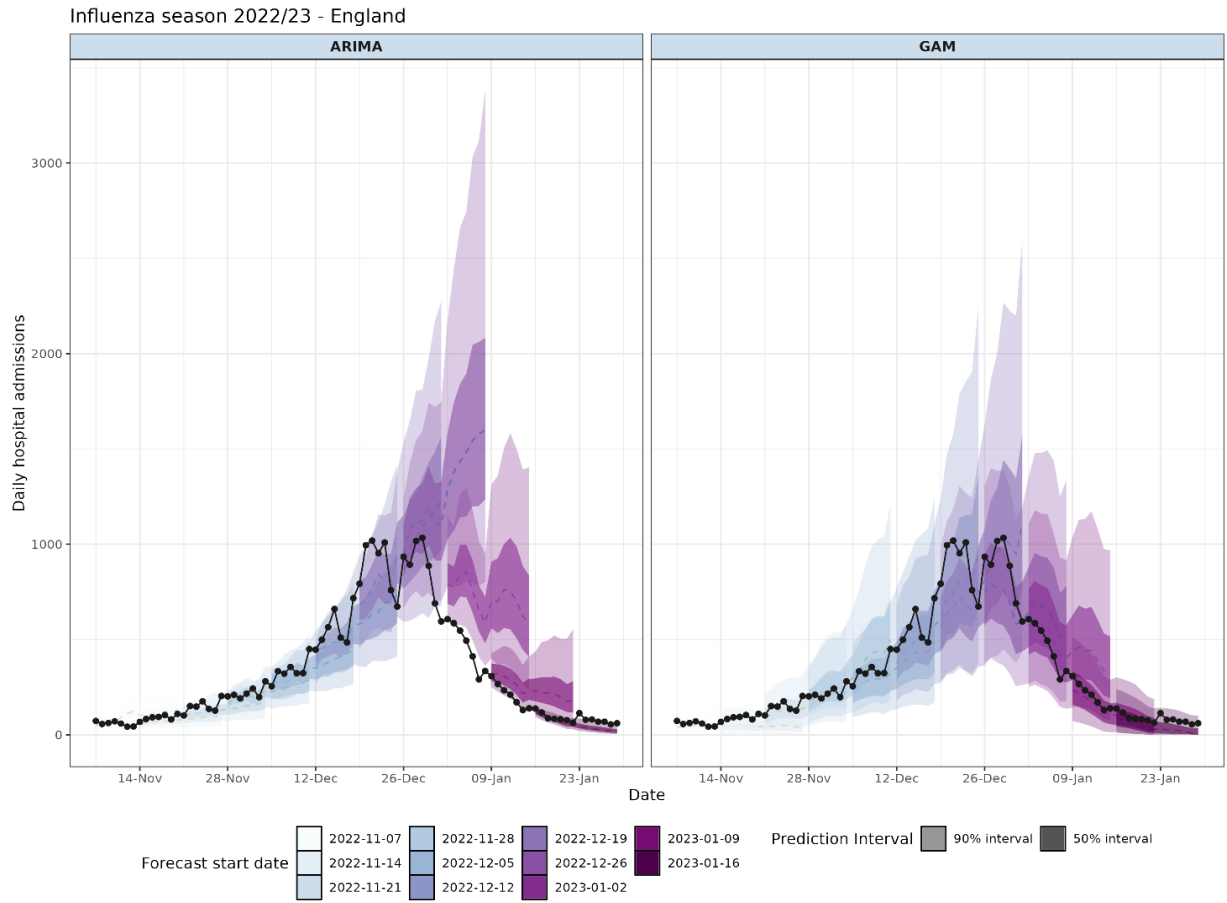

*Figure 2. Influenza season and model projections for England. Weekly forecasts of 14 days for the ARIMA and GAM models are shown throughout the influenza season. Black dots and lines represent the true admissions for each geographic breakdown.*

This picture of improved forecasts from the GAM continues when examining the region level forecasts (Figure 3). In multiple regions the ARIMA struggles to adapt to the turning around at the peak, such as in the South West, where consecutive forecast intervals are non-overlapping, though the GAM does not downturn well in the week commencing 02 Jan 2023. For some regions, such as the South East, the ARIMA model has substantially wider prediction intervals. From 09 Jan 2023 onwards, the GAM can detect a decline and reduce the width of the prediction intervals, however the ARIMA struggles to do this at a regional level, instead predicting 90% intervals that include growth, shown clearly in the Midlands. The models are fit to data; however, this data is of varying quality – as shown by the low counts observed in London and the East of England which are indicative of low ascertainment rates or poor reporting. The hierarchical component of the GAM allows for the national smoother to inform forecasts, pooling the trend from all STPs which helps improve performance.

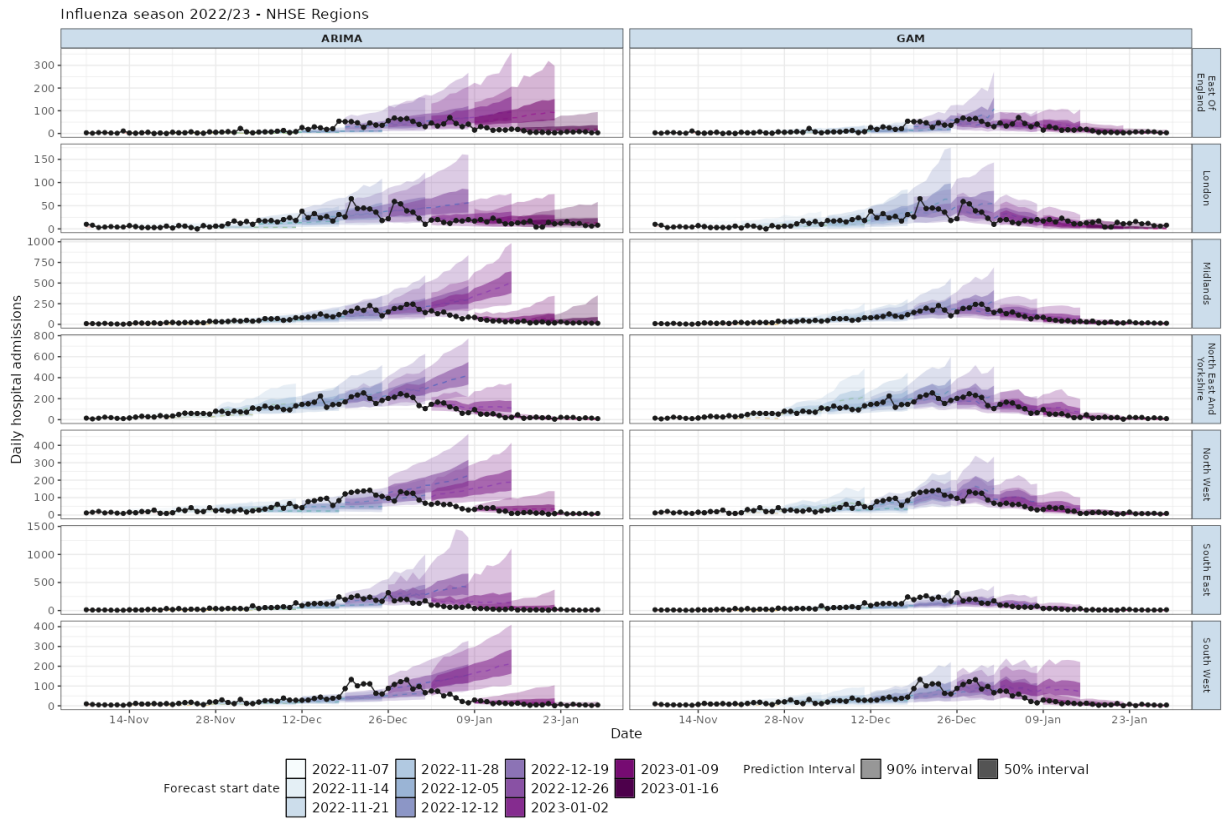

*Figure 3. Influenza season and model projections for England by commissioning region. Weekly forecasts of 14 days for the ARIMA and GAM models are shown throughout the influenza season. Black dots and lines represent the true admissions for each geographic breakdown. Note that y-axis scales differ between regions.*

### 3.3 Comparative performance

### 3.3.1 Performance over time

The out-of-sample performance for the interval score, MAE and bias are shown as the epidemic wave progresses through time for each model in Figure 4. Across most weekly forecasts, the GAM's bias is closer to zero than the ARIMA, which underpredicts during the epidemic growth phase then switches to overprediction at the peak and beyond. The MAE and interval score of each model is broadly consistent during the epidemic growth phase, but the scores diverge sharply at the epidemic peak, as shown in Figure 2 where the ARIMA overpredicts the peak and then struggles to decline fast enough. The interval score in this case is the preferred measure of performance to account for sharpness and under/overprediction, whilst MAE is in the natural units of the raw data.

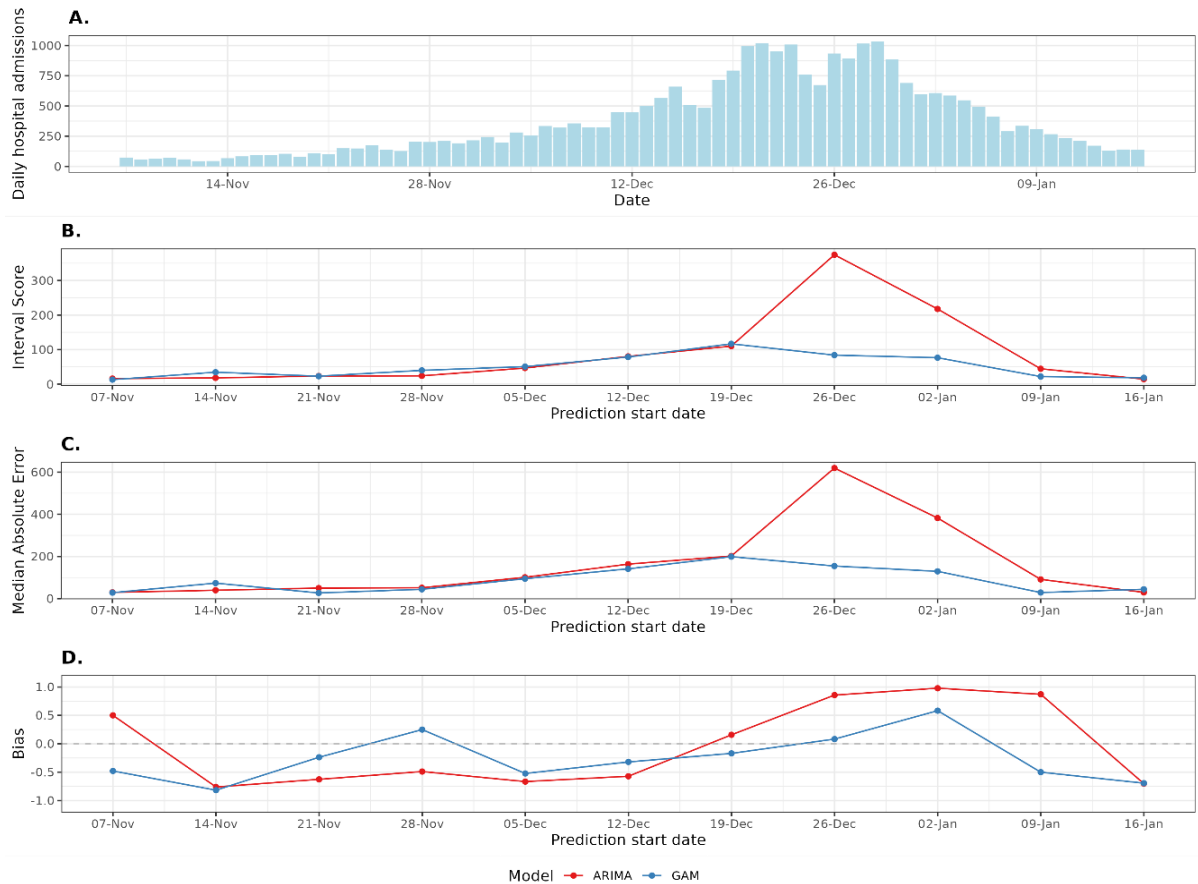

*Figure 4. The epidemic wave and corresponding GAM and ARIMA model performances for a forecast on the week start date. (A.) the counts of daily hospital admissions within the English 2022/23 winter season. (B.) Interval score for each week of forecasts showing the calibration over time. (C.) Median Absolute Error for each week of forecasts showing the error over time. (D.) Bias for each week of forecasts showing the under/overprediction over time.*

### *3.3.2 Forecast Horizon and Geography*

The hierarchical GAM introduced in this study consistently outperforms the ARIMA model across a wide range of scoring metrics, different geographies, across the whole forecast horizon, and for both $t_{max} + 7$ and $t_{max} + 14$, shown in Table 1. When presenting to decision makers, the national and regional forecasts are of most interest – where the GAM outperforms the ARIMA with lower MAE and better calibration (interval score). Understanding whether the models are likely to over or underpredict the admissions is also important when disseminating the forecasts to decision makers – for both metrics the GAM scores better than the ARIMA at coarse geographies, though the ARIMA has a better underprediction score at STP level. The MAE and interval scores of the ARIMA at the 7-day forecast horizon are comparable to the value of the GAM at 14 days, particularly at national and regional levels, indicating how we can have higher confidence in these forecasts. Across both 50% and 90% interval coverages the GAM is well calibrated compared to the ARIMA, giving us confidence that it will capture true values approximately in-line with expectations. The ARIMA on the other hand clearly fails, with some coverage metrics giving 0.00 indicating poor calibration, though the GAM has intervals clearly too wide for the 7-day horizon national aggregation.

| | | Evaluation scores for ARIMA and GAM models | | | | | |
|---|---|---|---|---|---|---|---|
| Model | Horizon | Interval Score | Underprediction | Overprediction | Median Absolute Error | 50% Coverage | 90% Coverage |
| National | | | | | | | |
| ARIMA | Overall | 88.1 | 12.7 | 47.6 | 160 | 0.279 | 0.773 |
| GAM | Overall | **50.6** | **9.83** | **4.56** | **87.8** | **0.669** | **0.981** |
| ARIMA | 7 | 71.1 | 8.40 | 39.7 | 146 | 0.000 | 0.727 |
| GAM | 7 | **39.0** | **6.39** | **1.63** | **58.5** | **0.727** | **1.000** |
| ARIMA | 14 | 169 | 21.0 | 105.0 | 294 | 0.182 | 0.545 |
| GAM | 14 | **82.3** | **13.8** | **18.70** | **163** | **0.455** | **0.909** |
| Regional | | | | | | | |
| ARIMA | Overall | 19.3 | 4.03 | 8.67 | 33.6 | 0.359 | 0.764 |
| GAM | Overall | **9.71** | **3.22** | **1.32** | **17.3** | **0.585** | **0.917** |
| ARIMA | 7 | 17.3 | **3.90** | 7.45 | 30.9 | 0.338 | 0.740 |
| GAM | 7 | **9.79** | 4.18 | **1.18** | **16.1** | **0.571** | **0.883** |
| ARIMA | 14 | 35.9 | 6.89 | 18.8 | 58.1 | 0.221 | 0.610 |
| GAM | 14 | **16.2** | **4.67** | **4.46** | **29.4** | **0.442** | **0.805** |
| STP | | | | | | | |
| ARIMA | Overall | 4.12 | **1.57** | 1.14 | 6.75 | 0.340 | 0.676 |
| GAM | Overall | **3.26** | 1.73 | **0.665** | **5.09** | **0.546** | **0.810** |
| ARIMA | 7 | 3.83 | **1.33** | 1.16 | 6.42 | 0.357 | 0.686 |
| GAM | 7 | **3.41** | 2.03 | **0.642** | **4.93** | **0.580** | **0.827** |
| ARIMA | 14 | 6.56 | **2.18** | 2.23 | 10.1 | 0.262 | 0.615 |
| GAM | 14 | **5.04** | 2.31 | **1.54** | **7.87** | **0.502** | **0.745** |

*Table 1. Interval score, underprediction, overprediction, absolute median error and coverage at 50% and 90% for the GAM and ARIMA models at national, regional and STP level. The Horizon column denotes how each model scores for the prediction at 7 and 14 days and across all predictions in the two-week forecast, from each week starting 07 Nov 2022 to 16 Jan 2023. The score in bold highlights the best score of the models for each horizon.*

### 3.4 GAM Spline Tuning

GAMs use splines to model smooth relationships between response and explanatory variables. The flexibility of splines to fit non-linearities in response variables is determined by the number of basis functions in each smoothing term, with more basis functions allowing for more non-linear relationships to be modelled, but at the increased risk of modelling noise as signal. The amount of smoothing is crucial to the model's predictive power; therefore, a sensitivity analysis was performed on the model's two main parameters. The performance of the GAM structure is shown in Figure 5 for multiple scoring metrics and a range of days per basis spline, for the national, $f_{nat}(t)$, and STP level $f_{stp_i}(t)$ smoothers. This shows the out-of-sample performance of the forecasts are highly dependent on $d_t$ for the national smoother (more variation along the x-axis), and to a lesser extent the STP smoother (less variation along the y-axis). The best performance of the national smoother occurs at 5 days per basis function consistently, though the best scoring for the STP smoother varies across metrics. As the bias metric is aggregated across the forecast horizon and epidemic wave the bias shown represents the average bias, which may be near zero due to both positive and negative bias over the period. The performance for all integer numbers of basis functions $t_d \in [1,7]$ are available in Supplementary Figure C to show more granularity nearer the optimal values.

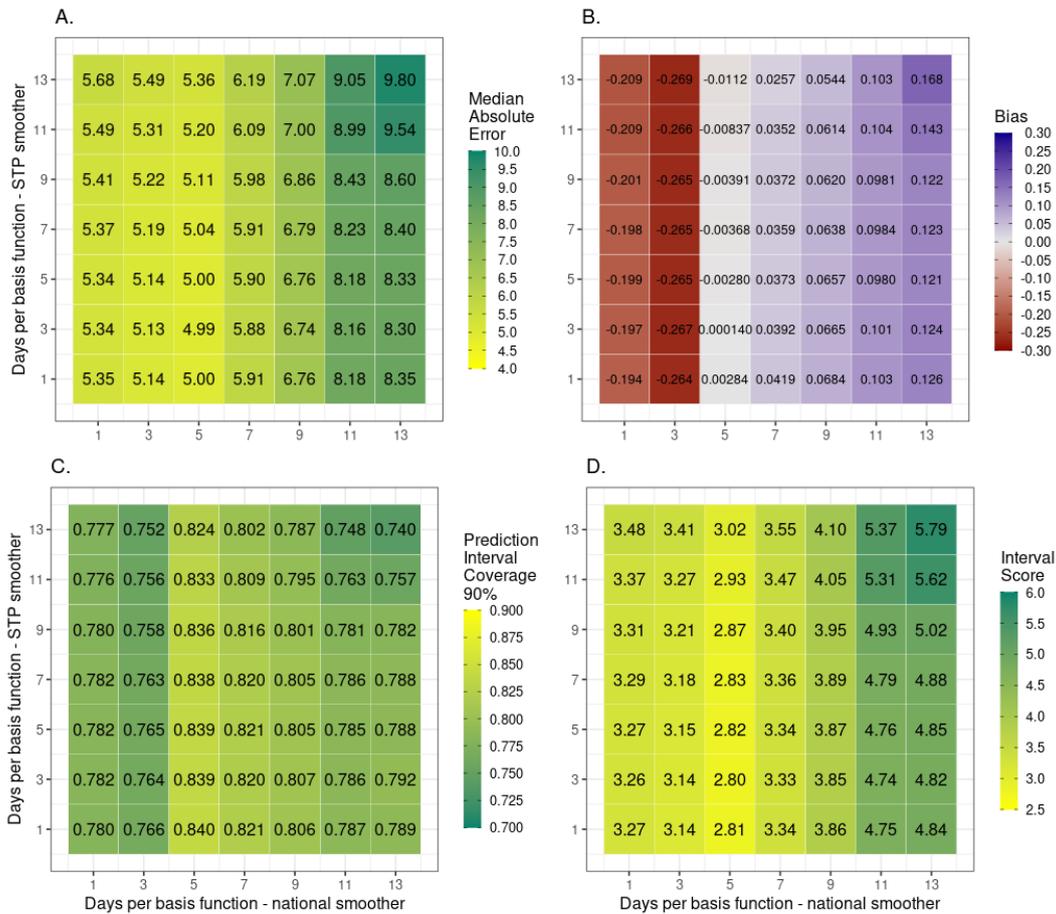

*Figure 5. The median absolute error (A.), bias (B.), prediction interval coverage (C.) and interval score (D.) metrics for combinations of different numbers of basis functions in the hierarchical GAM. Models were evaluated on held out forecasting performance for two-week projections, each week from 10 Nov 2022 to 08 Jan 2023.*

## 4. Discussion

### 4.1 Overview

In this study we show that a hierarchical GAM is an effective approach to short-term forecasting of influenza hospitalisations across England, outperforming the commonly used ARIMA approach. The model performs well both at low spatial levels (STP) and higher (region and national) aggregations. The GAM is impacted less by the expected issue of an autoregressive model at an epidemic peak, is less biased over time, and has a lower interval score than the ARIMA at both 7 and 14 days into the future. The model is sensitive to the number of basis functions in a smoother, particularly in the national component of the multi-level model, highlighting the need for careful parameter tuning.

### 4.2 Strengths

The hierarchical GAM introduced in this study has several strengths, making it a useful addition to the research in forecasting healthcare burdens due to infectious diseases. This modelling approach is easily applicable to other illnesses, does not rely on strong assumptions based on past seasonal patterns or epidemiological parameters and is robust when data quality may be low. Mechanistic models of influenza transmission and hospitalisation rely on detailed parametrisation (e.g. contact rates, infection hospitalisation rate, incubation and infectious period) and particular initial conditions (e.g. susceptible population, current prevalence). Exploring the pairwise combinations of different parameters and scenarios is computationally costly and contentious, as those providing model evidence during the COVID-19 pandemic can attest [40]. By taking a data-driven semi-mechanistic approach, we principally depend on subtle signals at fine spatial scales to detect upcoming changes in rates, and the computational efficiency of our method allows for weekly updates to projections.

By limiting the forecast horizon to 14-days this model avoids overextrapolation while still having high predictive performance and meeting the needs of policy makers. The spatial structure of the model, correlating adjacent STPs, and pooling the temporal trends through the hierarchical structure allows this model to detect the slowing growth rate near the peak of a wave taking advantage of the spatial variation in influenza peak times [41]. The explicit encoding of a nested random effect for the day-of-week effect allows the capturing of reporting and presentation patterns which the ARIMA model struggles to detect without a hierarchical component.

Evaluation of a probabilistic forecast is crucial to the operational use of a model. We employed proper scoring rules [42], particularly the interval score, to show the model calibration and out-of-sample performance in an unbiased and consistent way for quantile forecasts [42] [43].

*4.3 Limitations*

While this study introduces a novel and clearly effective technique in forecasting infectious disease hospitalisations – there are challenges associated with the approach. The reliance on fast turnaround granular SitReps are crucial to short term forecasting, however these can often have data quality issues requiring ad hoc fixes. Without these corrections a spline-based model may over extrapolate giving incorrect predictions due to the sensitivity of the splines. When deploying the novel method at the beginning of a season it cannot be known what the optimal parameter values over the whole wave will be, though consistent scoring with each new week of data will help mitigate this. What we cannot know until the next influenza season is whether this method performed well because of the pronounced epidemic shape of the influenza wave, which presented a clear growth then decay phase. Forecasting a season with a longer flatter plateau could be more challenging for this method, but the spatial component should still provide improvement on previous methods. The prediction intervals of this method are wide, which does introduce challenges for policy decisions, though this is also the case for the ARIMA model. Furthermore, the evaluation of this approach could be improved by incorporating the exponential structure into the scoring metrics [44]. In addition, the inclusion of leading indicators such as community testing, age structures, or syndromic surveillance could improve the predictive performance and confidence of the model, though this does increase the complexity of the model and number of dependencies required for real-time forecasting.

## 5. Conclusion

In this study we have introduced a novel method for short-term forecasting of influenza admissions waves, applied in real time during the England 2022/23 influenza season. The hierarchical structure, spatial components, and spline-based approach to capturing epidemic trends were shown to significantly out-perform as standard time-series approach across a range of scoring metrics, indicating a predictive performance at 14-days equivalent to the ARIMA performance at 7-days. The hierarchical GAM shown is simple to apply relying only on past time series and open lookup data, without strong assumptions on population immunity levels, vaccinations, or disease strains. The hierarchical and spatial component of the model allows the model to perform well at peak times comparatively to a pure autoregressive model. Further study into age structures of disease and leading indicators would perhaps improve model confidence, while coming at the cost of simplicity.

**Conflict of Interest**

The authors have declared that no competing interests exist. The authors were employed by the UKHSA but received no specific funding for this study.

**Data Availability Statement**

UKHSA operates a robust governance process for applying to access protected data that considers:

- the benefits and risks of how the data will be used
- compliance with policy, regulatory and ethical obligations
- data minimisation
- how the confidentiality, integrity, and availability will be maintained
- retention, archival, and disposal requirements
- best practice for protecting data, including the application of 'privacy by design and by default', emerging privacy conserving technologies and contractual controls

Access to protected data is always strictly controlled using legally binding data sharing contracts.

UKHSA welcomes data applications from organisations looking to use protected data for public health purposes.

To request an application pack or discuss a request for UKHSA data you would like to submit, contact [DataAccess@ukhsa.gov.uk](mailto:DataAccess@ukhsa.gov.uk).